\documentclass[12pt]{article}
\usepackage{color}   %May be necessary if you want to color links
\usepackage{hyperref}
\hypersetup{linktocpage}
\usepackage{pdflscape}
\usepackage[textwidth=15.2cm,textheight=22cm]{geometry}
\usepackage{amsmath,amssymb}
\usepackage{latexsym}
\usepackage{multicol}
\usepackage{graphicx}
\usepackage{caption}
\usepackage{cancel}
\usepackage{bbm}
\usepackage{bm}
\usepackage{cite}
\numberwithin{equation}{section}
\pdfoutput=1
\usepackage[T1]{fontenc}
\usepackage[latin9]{inputenc}
\usepackage[active]{srcltx}
\usepackage{esint}
\usepackage{cancel}
\usepackage{color}
\usepackage{ulem}
\usepackage{mathtools}
\usepackage{caption}
\allowdisplaybreaks[1]

\usepackage[symbol]{footmisc}

\renewcommand{\thefootnote}{\fnsymbol{footnote}}

\newcommand{\del}{\partial}
\newcommand{\be}{\begin{equation}}
\newcommand{\ee}{\end{equation}}
\newcommand{\bea}{\begin{eqnarray}}
\newcommand{\eea}{\end{eqnarray}}

\newcommand{\nn}{\nonumber}

\newcommand{\ndt}{\noindent}

\definecolor{ggreen}{rgb}{0.0, 0.5, 0.20}

\usepackage{tikz}
\usetikzlibrary{trees}
\usetikzlibrary{calc}
\usetikzlibrary {arrows.meta} 
\usepackage{lipsum}
\usepackage{varwidth}
\usepackage{pifont}
\usepackage{cancel}
\newcommand{\xdownarrow}[1]{%
  {\left\downarrow\vbox to #1{}\right.\kern-\nulldelimiterspace}
}

\usetikzlibrary{shapes.geometric, arrows, shadows}

\tikzstyle{arrow} = [thick,->,>=stealth]
\tikzstyle{arrow1} = [thick,<->,>=stealth]
\tikzstyle{line} = [draw, -latex']
\tikzset{
  double arrow/.style args={#1 colored by #2 and #3}{ thick,
    -latex,  line width=1.1*(#1),#2, % first arrow
    postaction={draw,-latex,#3,line width=1.9*(#1/2),
                shorten <=0.7*(#1)/3,shorten >=(#1)/3}, % second arrow
  }
}

\usepackage{tikz}
\usetikzlibrary{trees}
\usepackage{varwidth}
\usepackage{pifont}
\usepackage{cancel}
\usetikzlibrary{shapes.geometric, arrows, shadows}

\tikzstyle{forces} = [rectangle, rounded corners, minimum width=2cm, minimum height=0.8cm,text centered, draw=black, fill=white]
\tikzstyle{spin} = [rectangle, rounded corners, minimum width=1.3cm, minimum height=0.8cm,text centered, draw=black, fill=white]
\tikzstyle{theory} = [ellipse, minimum width=2.4cm, minimum height=0.8cm,text centered, draw=black, fill=white]
\tikzstyle{arrow} = [thick,->,>=stealth]
\tikzstyle{arrow1} = [thick,<->,>=stealth]
\tikzstyle{line} = [draw, -latex']
\tikzset{
	double arrow/.style args={#1 colored by #2 and #3}{
		-stealth,line width=#1,#2, % first arrow
		postaction={draw,-stealth,#3,line width=(#1/2),
			shorten <=(#1)/3,shorten >=2*(#1)/3}, % second arrow
	}
}

\usetikzlibrary{external}

\begin{document}

	\begin{titlepage}
		\begin{flushright}    
			{\small $\,$}
		\end{flushright}
		\vskip 1cm
		\centerline{\large{\bf{BMS symmetry in gravity: Front form versus Instant form}}}
		\vskip 1.5cm

		\centerline{Sudarshan Ananth$^\dagger$ and Sucheta Majumdar$^\ddagger$}%\footnote{corresponding author}}
		\vskip 1cm
		\centerline{$\dagger$\,\it {Indian Institute of Science Education and Research}}
		\centerline{\it {Pune 411008, India}}
		\vskip 1cm           
		\centerline{$\ddagger$\,\it{ENS de Lyon, CNRS, Laboratoire de physique,}}
			\centerline{\it{UMR 5672, F-69342 Lyon, France}}
\vskip 0.5cm
%\centerline{\textit{Email:} ananth@iiserpune.ac.in, sucheta.majumdar@ens-lyon.fr}
\vskip 1.5cm
		\centerline{\bf {Abstract}}
		\vskip 0.3cm
\ndt		In General Relativity, the allowed set of diffeomorphisms or gauge transformations at asymptotic infinity forms the BMS group, an infinite-dimensional extension of the Poincar\'e group. We focus on the structure of the BMS group in two distinct forms of Hamiltonian dynamics - the instant and front forms. Both similarities and differences in these two forms are examined while emphasising the role of non-covariant approaches to symmetries in gravity.
\vskip 2.5cm
\centerline{\textit{This article is dedicated to the memory of Lars Brink}}
	\vskip 0.7cm

	\vskip 0.5cm
\centerline{Essay written for the}
\centerline{Gravity Research Foundation 2023 Awards for Essays on Gravitation}
\vskip 0.5cm

		\vfill
	\end{titlepage}
\renewcommand*{\thefootnote}{\arabic{footnote}}

\ndt
Symmetries, both spacetime and gauge, serve as powerful tools in formulating consistent physical theories. Symmetries are essential signposts when exploring theories like general relativity which lacks a consistent quantum mechanical description. Gauge symmetries, which have to do with the redundancies in our description of a physical theory, offer multiple perspectives on the same theory. 
\vskip 0.3cm
\ndt Asymptotic symmetries are a class of spacetime symmetries that appear as one approaches infinity.  One may move along a space-like direction to approach spatial infinity, or along a null direction - towards null infinity. At asymptotic infinity, the symmetry group of gravity, known as the Bondi-van der Burg-Metzner-Sachs (BMS) group~\cite{Bondi:1962px, Sachs}, is larger than the na\"ively-expected Poincar\'e group. The BMS symmetry involves an infinity of parameters, serving as a generalization of the finite-dimensional Poincar\'e group. The BMS symmetry sets general relativity, where the spacetime metric is dynamical, apart from special relativity where the asymptotic symmetry group remains Poincar\'e.
\vskip 0.3cm
\ndt To describe the evolution of a relativistic system in time, Dirac proposed three distinct forms of Hamiltonian dynamics~\cite{Dirac}. The forms vary essentially in their choice of the time coordinate or `evolution parameter'. The light-cone form involves two null coordinates $x^\pm =(t\pm x^3)/ \sqrt 2$ along the light-cone and two transverse spatial directions, resulting in a $(2+2)$ split of four-dimensional spacetime. The dynamics in this frame, also called the \textit{front form}, is clearly in contrast with the standard description where {\it {one}} time coordinate is singled out (the dynamics is then described on a constant time hypersurface spanned by three spatial directions). Thus this approach prefers a $(3+1)$ split of spacetime and is also referred to as the \textit{instant form}. Another obvious candidate when choosing a time coordinate would be the proper time referred to as the \textit{point form} (from the study of the relativistic point particle). This article focusses on the front and instant forms of dynamics, since these forms have more relevance to the study of massless particles. Starting with the structure of the Poincar\'e group in the instant and front forms, we discuss how the enhancement to the infinite-dimensional BMS group occurs in both these cases, for theories of gravity.
\vskip 0.3cm
\ndt  A clear picture of the symmetries in a system in one or more of these forms of Hamiltonian dynamics is essential for the subsequent quantization procedures. Symmetries in the Hamiltonian formulation are realized through canonical generators expressed in terms of phase space variables, a set of physical fields $\{\Phi\}$ and their conjugate momenta $\{\pi_{\Phi}\}$. The phase space is endowed with a Poisson bracket $[\Phi(x), \pi_\Phi(y)] \sim \delta (x-y)$. A canonical transformation $X$ in the phase space corresponds to a symmetry if it leaves the Hamiltonian action $\mathcal S_H$ invariant 
	\[ (\Phi, \pi_\Phi) \xrightarrow{\text{\quad X\quad }} (\tilde{\Phi}, \tilde{\pi}_\Phi)\, \quad \text{such that}\quad \delta_X \mathcal S_H = 0\,.\]
	 The transformation law of the fields may be derived from the canonical generator $G[X]$ through the Poisson bracket
	\[ \delta_X \Phi= [ G[X], \Phi ] \, ,\quad \delta_\xi \pi_\Phi= [ G[X], \pi_\Phi ] \,.\]
	In the quantum theory, using the canonical quantization scheme, the Poisson bracket is promoted to the commutator, $[\ ,\ ]_{PB} \rightarrow i\hbar [\ ,\ ]$. The conserved quantities associated with the symmetries play the role of physical observables in the quantum theory.
	
	\vskip 0.3cm
\ndt
 Quantization of gauge theories is, however, a particularly delicate issue because the Hamiltonian description in these theories involve constraints. The constraints imply that not all phase space variables are dynamical, which reflects the gauge redundancy in the theory. In the instant form, such constrained Hamiltonian systems can be studied using the Dirac-Bergmann algorithm~\cite{Dirac:1950pj, Bergmann:1949zz} to classify the constraints and analyze their algebra in the context of symmetries. In the front form, a key advantage is that the constraints may be solved completely in order to eliminate the gauge degrees of freedom. This allows us to obtain a Hamiltonian formulation solely in terms of the physical fields of theory.
\vskip 0.3cm
\ndt	
\textit{Structure of Poincar\'e algebra}

	\vskip 0.3cm
	\ndt 
For quantum field theories in flat spacetime, the spacetime symmetry is parameterized by $\xi^\mu = \omega^\mu{}_\nu x^\nu + a^\mu$, corresponding to the Poincar\'e group. The \textit{constant} parameters, $a^\mu$ and $\omega^\mu{}_\nu$, label spacetime translations and Lorentz rotations respectively. In terms of structure, there are two natural ways to divide up this algebra.\\

	\ndt 1. $P-M$ split: $G[\xi] = a^\mu P_\mu + \omega^{\mu \nu} M_{\mu \nu} $
	 \vskip 0.2cm
    	 
\ndt	 The four spacetime translations are generated by $P^\mu$ while the $M^{\mu\nu}$ produce the Lorentz transformations. The Poincar\'e algebra in this split has the following structure
	 \be \label{P-M}
	    [M,M] \sim M\,, \quad [P, M] \sim P\,, \quad [P, P] \sim 0\,.
	  \ee
This split is ideal for working with theories formulated in a covariant manner. \\	
	    
\ndt 2. $\mathcal K-\mathcal D$ split:  $G[\xi] = \xi^{\text{space}} \mathcal K + \xi^{\text{time}} \mathcal D $
	  	 \vskip 0.2cm

\begin{figure}[h]
\centering
\begin{tikzpicture}[dot/.style={circle,inner sep=1pt,fill,label={#1},name=#1},
  extended line/.style={shorten >=-#1,shorten <=-#1},
  extended line/.default=1cm]
\draw [line width = 0.2mm] (-3.5,-0.7) -- (3.5,-0.7);
\draw [line width = 0.2mm] (-2,0.7) -- (5,0.7  );
\draw [line width = 0.2mm] (-3.5,-0.7) -- (-2, 0.7);
\draw [line width = 0.2mm] (3.5,-0.7) -- (5,0.7);
\draw[-{Stealth[length=3mm]}, dashed, line width = 0.3mm] (0.7,0.0) -- (0.7,3) node[above] {$x^0 =t$};
\draw[-{Stealth[length=3mm]}, dashed, line width = 0.3mm] (0.7,-0.7) --(0.7,-3);
\draw[-{Stealth[length=3mm]}, line width = 0.4mm] (0.7,0.0) -- (0.7,1.0) ;
\draw[-{Stealth[length=3mm]}, dashed, line width = 0.3mm] (0.7,0) -- (-4,0); 
\draw[-{Stealth[length=3mm]}, dashed, line width = 0.3mm] (0.7,0) -- (5.5,0) node[above,xshift=2mm] {$x^1$};
\draw[-{Stealth[length=3mm]}, line width = 0.4mm] (0.7,0.0) -- (1.7,0) ;
\draw[-{Stealth[length=3mm]}, dashed, line width = 0.3mm] (0.7,0) --(-0.7, -1.5);
\draw[-{Stealth[length=3mm]}, dashed, line width = 0.3mm] (0.7,0) -- (2.2,1.4) node[right, yshift= 2mm]{$\{x^{2,3}\}$} ;
\draw [-{Stealth[length=2mm]}, line width = 0.3mm] (4.5,0.5) .. controls (4.6, 1.2) and (5.2, 1.4).. (5.9, 1.1) node[right, xshift = 0mm]{$\Sigma: t=0$};
\end{tikzpicture}
\caption{The $t=0$ spatial surface in instant form labelled by coordinates $(x^1,x^2, x^3)$ } \label{fig:instant}
\end{figure}
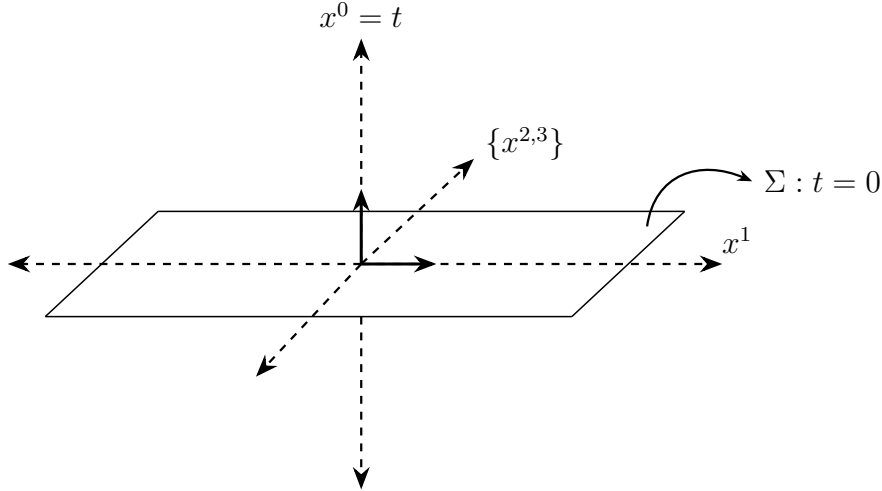

\ndt The time component of $\xi^\mu$ corresponds to timelike diffeomorphisms, namely time translation and Lorentz boosts. On the other hand, the spatial components are associated with the transformations along a constant time hypersurface that includes spatial translations and rotations. So the kinematical part of the algebra $\mathcal K $ acts on a constant time surface, while the dynamical part $\mathcal D$ maps the initial hypersurface to another hypersurface at a later time. The algebra structure takes the form
		  \be \label{K-D}
	    [\mathcal K,\mathcal K] \sim \mathcal K\,, \quad [\mathcal D, \mathcal K] \sim \mathcal D\,, \quad [\mathcal D, \mathcal D] \sim 0\,.
	  \ee 
	 The dynamical generators collectively play the role of the Hamiltonian operator in quantum field theory, unlike in quantum mechanics where the time derivative $\del_t$ alone suffices.
	 \vskip 0.3cm
\ndt
In the instant form, we have the time $x^0=t$ and three spatial coordinates $x^i$, labeled by a constant time hypersurface, as depicted in figure \ref{fig:instant}.

\ndt Thus, the Kinematical generators in instant form are the three spatial translations $P^i$ and three Lorentz rotations $J^i$ . The dynamical objects consist of the time translation generator $P^0$ and three Lorentz boosts $B^i$, resulting in the four Hamiltonians
\[  \mathcal K \equiv \{P^i, J^i = \epsilon^{ijk}M_{jk}\}\,, \quad  \mathcal D \equiv \{P^0, B^i= M^{0i}\}\, , \quad i, j, \ldots = 1, 2, 3\,.\]	

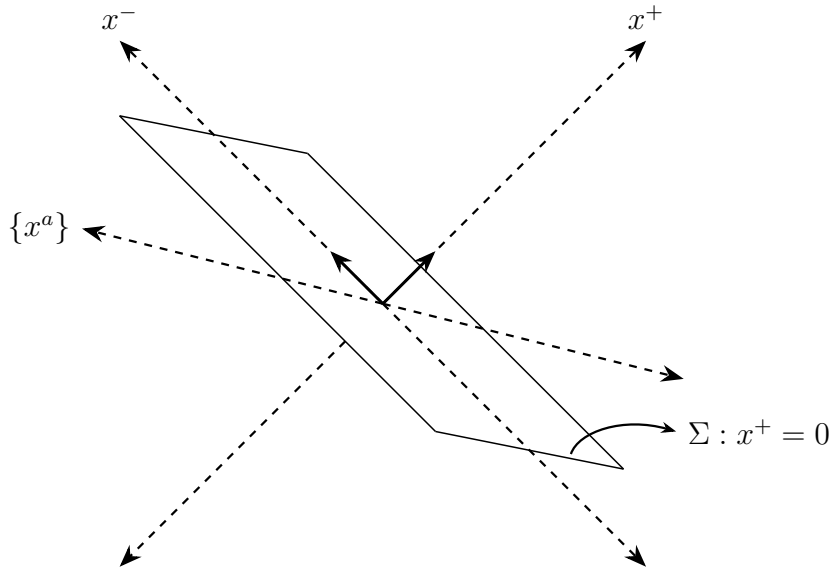
\begin{figure}[h]
\centering
\begin{tikzpicture}[dot/.style={circle,inner sep=1pt,fill,label={#1},name=#1},
  extended line/.style={shorten >=-#1,shorten <=-#1},
  extended line/.default=1cm]
\draw [line width = 0.2mm] (-1,2) -- (-3.5,2.5);
\draw [line width = 0.2mm] (0.7,-1.7) -- (3.2,-2.2);
\draw [line width = 0.2mm] (0.7,-1.7) -- (-3.5, 2.5);
\draw [line width = 0.2mm] (-1,2) -- (3.2,-2.2);
\draw[-{Stealth[length=3mm]}, dashed, line width = 0.3mm] (0.0,0.0) -- (3.5,3.5) node[above] {$x^+$};
\draw[-{Stealth[length=3mm]}, dashed, line width = 0.3mm] (-0.5,-0.5) --(-3.5,-3.5);
\draw[-{Stealth[length=3mm]}, line width = 0.4mm] (0.0,0.0) -- (0.7,0.7) ;
\draw[-{Stealth[length=3mm]}, dashed, line width = 0.3mm] (0,0) -- (3.5,-3.5); 
\draw[-{Stealth[length=3mm]}, dashed, line width = 0.3mm] (0,0) -- (-3.5,3.5) node[above] {$x^-$};
\draw[-{Stealth[length=3mm]}, line width = 0.4mm] (0.0,0.0) -- (-0.7,0.7) ;
\draw[-{Stealth[length=3mm]}, dashed, line width = 0.3mm] (0,0) --(4, -1);
\draw[-{Stealth[length=3mm]}, dashed, line width = 0.3mm] (0,0) -- (-4,1) node[left]{$\{x^a\}$};
%\draw[-{Stealth[length=3mm]}, line width = 0.4mm] (0.0,0.0) -- (-1,0.25) ;
%\draw[-{Stealth[length=2mm]},  line width = 0.2mm] (-1.3,2.1) ..controls (-1.0, 2.3)  and (-0.7, 1.9).. (-0.3,2.7) ;
%\draw[-{Stealth[length=2mm]},  line width = 0.2mm] (1.3,- 2.1) ..controls (1.0, -2.3)  and (0.7, -1.9).. (0.3,-2.7) ;
\draw [-{Stealth[length=2mm]}, line width = 0.3mm] (2.5,-2.0) .. controls (2.6, -1.7) and (3.2, -1.5).. (3.9, -1.7) node[right, xshift = 0mm]{$\Sigma: x^+=0$};
\end{tikzpicture}
\caption{The $x^+=0$ null surface in front form labelled by coordinates $(x^-, x, \bar x)$} \label{fig:null}
\end{figure}

\ndt The front form involves two null coordinates $x^{\pm}$, of which either may be chosen to be the time coordinate, and two transverse spatial coordinates. Choosing $x^+$ to be light-cone time, the initial constant time hypersurface is the $x^+ = 0$ surface. The coordinates on this three-dimensional hypersurface are the other null coordinate $x^-$ and two transverse complex coordinates, denoted by $x^a$ here, as shown in figure~\ref{fig:null}.

\vskip 0.3cm
\ndt 
The kinematical part of the algebra in the front form involves the three translation generators, $(P^+, P^a)$ and four rotation generators $(J^{ab}, J^{+a}, J^{+-})$. The dynamical generators are $P^-$ and $J^{-a}$
\[ \mathcal K \equiv \{P^+, P^a, J^{ab}, J^{+a}, J^{+-} \}\, , \quad \mathcal D \equiv \{P^-, J^{-a} \}\,, \quad a, b, \ldots = x, \bar x\,. \]
On the $x^+= 0$ surface, the $J^{-+}$ is kinematical.

\ndt Three Hamiltonians, as opposed to four in the instant form, are easier to work with, in deriving interacting Hamiltonians~\cite{BBB}. Unlike in the usual case, Lorentz boosts in the light-cone Poincar\'e algebra commute with each other. This highlights the underlying Carrollian structure of the light-cone Poincar\'e group~\cite{Bacry:1968zf}. 
	\vskip 0.3cm
	\ndt
	\textit{From Poincar\'e to BMS}
	\vskip 0.3cm
	\ndt
	Diffeomorphism invariance that lies at the heart of general relativity allows for a change of coordinates $x^\mu \rightarrow x^\mu +\xi^\mu (x)$ with an arbitrary local parameter $\xi^\mu$. This \textit{gauge} freedom in gravity is associated with the Hamiltonian and momentum constraints in the Dirac-ADM formulation of gravity.  As alluded to before, the asymptotic symmetry group of gravity is the BMS group, which comprises Lorentz transformations and infinitely many supertranslations. The study of asymptotic symmetries is very sensitive to the choice of coordinate system, gauge and boundary conditions imposed on the fields. This is because the asymptotic symmetries are nothing but residual gauge transformations or diffeomorphisms that do not vanish at the boundary of spacetime. There are multiple approaches to the study of asymptotic symmetries, including the following three key ones.
		 
	 \begin{itemize}
\item Bondi Aproach: This approach employs a special set of coordinates, known as the Bondi coordinates, which in four dimensions involve one null time coordinate, one radial coordinate and two spatial coordinates. The BMS symmetry in gravity was originally discovered at null infinity in this setup~\cite{Bondi:1962px, Sachs}. For asymptotically flat spacetimes, the BMS symmetry arises at null infinity as the set of allowed diffeomorphisms that preserve the form of the metric at the asymptotic boundary. The BMS group at null infinity has been studied extensively and further enhanced to include superrotations, Diff($\mathbb S^2$) and other larger groups.
	\item Carrollian approach: The Carroll group has been formally known to exist in the literature as the `ultrarelativistic' $c\rightarrow$ limit of the Poincar\'e group for decades~\cite{Bacry:1968zf}, but had not been associated with any physical setup until recently. In 2014, the conformal extension of the Carroll group was shown to be isomorphic to the BMS group~\cite{Duval:2014uva}. This led to a rebirth of the Carroll group as the spacetime symmetry of null hypersurfaces, that are formed by a plane wave propagating with the velocity of light. This also resulted in new efforts to formulate ultrarelativistic theories with the Carroll group as the underlying symmetry as opposed to relativistic theories with Lorentz symmetry.
	\item Hamiltonian approach: The BMS symmetry can also be realised using Hamiltonian methods in one of more forms of relativistic dynamics discussed above. This serves as a vital step forward from a formal field-theoretic standpoint as it offers a consistent canonical description of BMS symmetry in terms of phase space variables, crucial for investigating implications for the quantum theory. We now discuss the canonical realization of BMS in the instant and front forms of Hamiltonian dynamics.
\end{itemize}
	\vskip 0.3cm
	\ndt
	\textit{BMS symmetry in instant form}
	\vskip 0.3cm
	\ndt
	In the (3+1) Dirac-ADM formulation~\cite{Dirac:1958sc, Arnowitt:1962hi} with standard boundary conditions associated with well-defined canonical charges, the asymptotic symmetry group was found to be just Poincar\'e and not BMS. This canonical realization of the Poincar\'e algebra in the phase space of gravity in the instant form can be found in~\cite{Regge:1974zd}. Attempts to recover the BMS group at spatial infinity were met with failure, as the required choice of boundary conditions would lead to ill-defined symmetry generators and conserved charges. This gap in our knowledge was bridged recently, where the BMS group was obtained at spatial infinity by virtue of meticulously relaxing the boundary conditions while respecting the notion of canonical conserved charges. 
The enhancement of the Poincar\'e group to BMS occurs in the translation part of the algebra. As a result of the relaxed boundary conditions introduced in~\cite{Henneaux:2018cst}, the spacetime translations $P^\mu$ in Poincar\'e labelled by the four constant parameter $a^\mu$ are replaced by an infinite set of supertranslations $\mathcal{ST}$ labelled by an angle-dependent or local parameter $\tau(\theta, \phi)$. Here, the angles $\theta$ and $\phi$ may be identified with the coordinates of the sphere at infinite radial distance $r$ as in the spherical polar coordinates $x^i = (r, \theta, \phi)$. The Lorentz part $M^{\mu\nu}$ of the algebra, however, remains unchanged. Thus, the map from the ten-dimensional Poincar\'e group to the infinite-dimensional BMS group in the instant form can be schematically written as
		\be  M^{\mu \nu} \rightarrow M^{\mu \nu}\,, \quad P^\mu\rightarrow \mathcal{ST}\,. \nn
		\ee
		The BMS algbera has the same form as in~\ref{P-M} with $P^\mu$ replaced with the supertranslations $\mathcal{ST}$. The local parameter $\tau(\theta, \phi)$ can be decomposed into even and odd modes in a basis of spherical harmonics $Y_{lm}(\theta, \phi)$
		\[ \tau(\theta, \phi) = T_{0,0} Y_{0,0} + \sum_{m=-1}^{1} T_{1,m} Y_{1,m} + \sum_{m=-2}^{2} T_{2,m} Y_{2,m}+ \cdots \,. \]
		The Poincar\'e subgroup of BMS can be obtained by the restricting the parameter to lowest four harmonics, where $T_{0,0}$ corresponds to the time translation and $T_{1,m}$ to the three spatial translations. The structure of the BMS group at spatial infinity, thus, matches with the original interpretation of BMS at null infinity, wherein angle-dependent supertranslations replace the four ordinary translations.
		\vskip 0.5cm
		
\begin{center}
* ~ * ~ *
\end{center}

%\vskip 0.5cm

	\subsubsection*{The need for a unifying framework}
There has been significant progress recently, in the study of asymptotic symmetries in the context of modern scattering amplitude techniques, such as soft theorems, double copy and twistor theory (see for example~\cite{Strominger}). BMS symmetries have been shown to constrain the gravitational scattering problem. But the two sides of this connection, namely asymptotic symmetries and scattering amplitudes, are conventionally studied in different setups. Asymptotic symmetries are typically examined as a boundary-value problem of classical solutions at null or spatial infinity while scattering amplitudes are investigated using on-shell approaches that often involve working in a physical gauge, such as the light-cone gauge~\cite{MHV light-cone, Ananth:2007zy}. 
\vskip 0.2cm
\ndt It is then natural to ask whether we can identify one framework that will offer a unified description of both asymptotic symmetries and scattering amplitudes. A light-cone realization of the BMS algebra constitutes a first step in that direction, one platform from which we may examine asymptotic symmetries from an action principle, which is also conducive for deriving on-shell physics. Instead of working with the boundary data available at asymptotic infinity, we can formulate an initial-value problem in the bulk spacetime, where we study the evolution of the initial data on a null hypersurface.
	\vskip 0.5cm
	\ndt 
	\textit{BMS symmetry in front form}
	\vskip 0.3cm
	\ndt In the front form, the gauge constraints can be solved to eliminate unphysical degrees of freedom. In the case of gravity, the light-cone action is written entirely in terms of two physical modes and contains only first-order time derivatives (an inherently Hamiltonian system). This (2+2) Hamiltonian formulation of gravity in the front from was first presented in~\cite{Scherk:1974zm} and the corresponding light-cone Poincar\'e algebra was studied in~\cite{BBB}. The light-cone BMS symmetry is again a local extension of the light-cone Poincar\'e algebra, but the difference from the instant form is that the enhancement occurs in a different part of the algebra~\cite{Ananth:2020ngt}. In light-cone BMS, the time component of the diffeomorphisms $\xi^+$ gets enhanced from a global to a local parameter, $\xi^+ = T(x, \bar x)$. Therefore, the extension from Poincar\'e to BMS algebra affects the dynamical part of the algebra. Instead of three dynamical tranformations $P^-$, $J^{-a}$ we end up with infinitely many `supertranslations' denoted by $\mathcal D(T)$. The kinematical part of the algebra remains the same and hence, the map from light-cone Poincar\'e to light-cone BMS is schematically
	\[ \mathcal K \rightarrow \mathcal K \,,\quad \mathcal D \rightarrow \mathcal D(T) \,,\]
	in contrast to the instant form case. The BMS algbera, then, has the same form as in~\ref{K-D} with $\mathcal D$ replaced by the supertranslations $\mathcal{D}(T)$. 
	\vskip 0.2cm
	\ndt The local parameter $T(x, \bar x)$ may be decomposed in a complex coordinate basis $(x, \bar x)$
\[	T(x,\bar x)= c_{0,0} + c_{1,0} x + c_{0,1} \bar x + c_{1,1} x\, \bar x + \cdots + c_{m,n} x^m \bar x^n\,. \nn \]

	\ndt The Poincar\'e subgroup of the light-cone BMS is then the restriction to the lowest three modes that are at most linear in coordinates, with $c_{0,0}$ corresponding to time translations and $c_{1,0}$, $c_{0,1}$ corresponding to the two boosts. Clearly, from a group theory perspective, the BMS structure here is very different from the instant form. Instead of the normal subgroup of Poincar\'e $P^\mu$, the enhancement in this case occurs in the Abelian ideal $\mathcal D$. In recent work, the BMS algebra was also obtained as the residual gauge symmetry in light-cone gravity without any explicit reference to asymptotic limits~~\cite{Ananth:2020ojp}. This result resonates with the idea that the BMS symmetry may be more than just an asymptotic symmetry of the theory~\cite{Duval:2014uva}. 
	
	\begin{figure} [h]
\centering
\includegraphics[width=13cm]{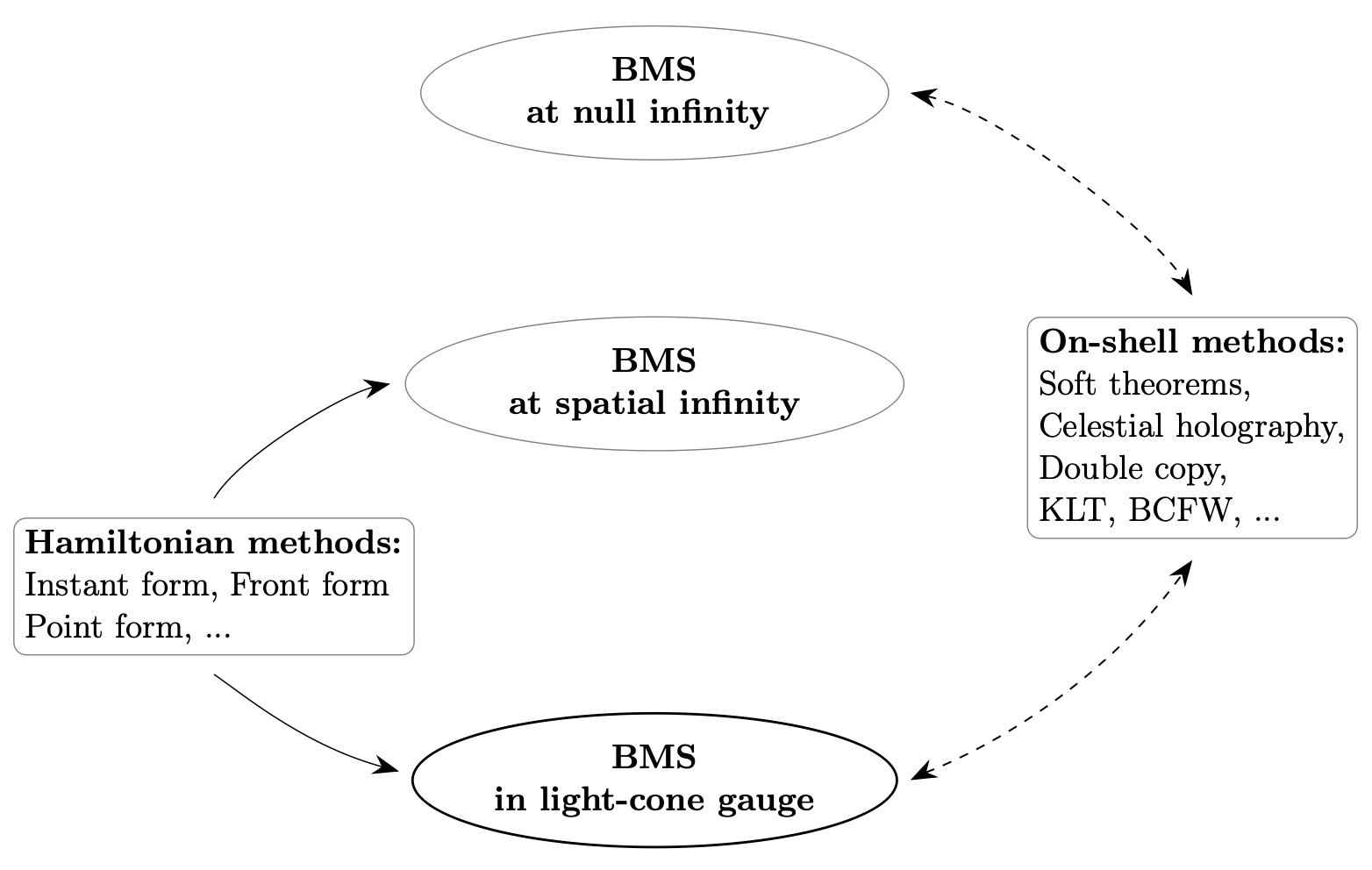}
		\caption{Light-cone BMS as a bridge between Hamiltonian and on-shell methods} \label{fig:schematic}
		\end{figure}

\ndt In the light-cone formulation, one can eliminate all unphysical gauge degrees of freedom from the theory, a feature that has proved to be fruitful in many instances. Therefore, it serves as a natural starting point for such a unifying framework because (i) one can reformulate the asymptotic symmetries as an initial-value problem in the front form and (ii) the transition to on-shell methods in a physical gauge is quite straightforward.  As a matter of fact, the light-front is a natural basis for scattering amplitudes with spinor helicity structures emerging naturally~\cite{Ananth:2012}. It is therefore worth exploring the potential links between the asymptotic symmetries in the front form and celestial or flat space holography constituting an effort to establish a duality between the S-matrix of four-dimensional quantum gravity and correlation functions in a two-dimensional ``celestial'' CFT, akin to the idea of AdS/CFT duality. 

\vskip 0.5cm
\ndt {\it {Acknowledgments}}
\vskip 0.1cm
\ndt The work of SA is partially supported by a MATRICS grant - MTR/2020/000073 - of SERB. The work of SM is supported by  the LABEX Lyon Institute of Origins (ANR-10-LABX-0066) Lyon within the Plan France 2030 of the French government operated by the National Research Agency (ANR).

\end{document}